\documentclass[times,letterpaper,10pt,twocolumn]{article} 
\usepackage{latex8}
\usepackage{iasted}
\usepackage{times}
\usepackage{epsfig}
\usepackage{latexsym}
\usepackage{amsmath}
\begin{document}
\newtheorem{definition}{Definition}[section]
\newtheorem{assum}{Assumption}[section]
\newenvironment{as}[1]{\begin{assum}[#1] \em}{\mbox{} \newline
                                                        $\Box$ \end{assum}}
\def\cost(#1){\mathify{\| #1 \|}}
\newtheorem{theorem}{Theorem}[section]
\newtheorem{lemma}{Lemma}[section]
\newtheorem{property}{Property}[section]
\newtheorem{corr}{Corollary}[section]
\newtheorem{conj}{Conjecture}[section]
\newtheorem{fact}{Fact}[section]
\newtheorem{ex}{Example}[section]
\newenvironment{vb}{\begin{ex} \em \mbox{} \newline}{\mbox{} \newline  \end{ex}}
\newenvironment{proof}{{\bf Proof:}}{\mbox{} {\hfill}
					$\Box$}
\newenvironment{alg}[1]{{\bf Algorithm (#1)} }{\mbox{}
\newline \hspace*{\fill}}

\date{}
\title{Efficient Algorithms and Routing Protocols for Handling Transient Single Node Failures}
\author{Amit M. Bhosle\thanks{Currently at Amazon.com, 1200 12$^{th}$ Ave.
S., Seattle, WA - 98144} $~~~$ and $~~\!$ Teofilo F. Gonzalez\\
\\
Department of Computer Science\\
University of California\\
Santa Barbara, CA 93106\\
\{bhosle,teo\}@cs.ucsb.edu
}
\maketitle
\begin{abstract}
{\em Single} node failures represent more than 85\% of all node
failures\cite{mibcd} in the today's large communication networks such as the
Internet. Also, these node failures are usually 
{\em transient}. Consequently, having the routing paths 
globally recomputed does not pay off
since the failed nodes recover fairly quickly, and the recomputed
routing paths need to be discarded. Instead, we develop algorithms
and protocols for dealing with such transient single node failures
by suppressing the failure (instead of advertising it across the
network), and routing messages to the destination via alternate paths 
that do not use the failed node. 
We compare our solution to that of \cite{znylwc}, which also
discusses such a {\em proactive recovery scheme} for handling
transient node failures. We show that our algorithms are faster 
by an order of magnitude while our paths are equally good.
We show via simulation results that our paths are usually
within 15\% of the optimal for randomly generated graph with 
100-1000 nodes. 
\end{abstract}

\section{Introduction}
Let $G=(V,E)$ be an edge weighted graph that
represents a computer network, where the weight (positive real number),
denoted by $cost(e)$,
of the edges
represents the cost (time) required to transmit a packet through
the edge (link).
The number of vertices ($|V|$) is $n$ and the number of
edges ($|E|$) is $m$.
It is well known that a shortest paths tree of a node $s$,
$\mathcal{T}_s$, specifies the
fastest way of transmitting a message to node $s$ originating
at any given node in the
graph under the assumption that
messages can be transmitted at the specified costs.
Under normal operation the routes are the fastest, but when the
system carries heavy traffic on some links these routes might not be
the best routes.
These trees can be constructed (in polynomial time) by finding
a shortest path between every pair of nodes.
In this paper we consider the case when the
nodes in the network are susceptible to transient faults.
These are sporadic faults of at most one node\footnote{The nodes are
{\em single-} or {\em multi-}processor computers} at a time
that last for a relatively short period of time.
This type of situation has been studied in the past 
\cite{znylwc} because it represents
most of the node failures occurring in networks.
{\em Single} node failures represent more than 85\% of all node
failures \cite{mibcd}. Also, these node failures are 
usually {\em transient},
with 46\% lasting less than a minute, and 86\% lasting less than
10 minutes \cite{mibcd}. 
Because nodes fail for relative short periods of time, propagating
information about the failure throughout the network is not recommended.

In this paper we consider the case where the network is {\em biconnected} 
({\em 2-node-connected}), meaning that
the deletion of a single node does not disconnect the network.
Based on our previous assumptions about failures, a message
originating at node $x$ with destination $s$ will be sent along the
path specified by $\mathcal{T}_s$ until it reaches node $s$ or
a node (other than $s$) that failed.
In the latter case, we need to use a
recovery path to $s$ from that point.
Since we assume single node faults and
the graph is biconnected, such a path always exists.  We call this
problem of finding the recovery paths the
{\em Single Node Failure Recovery (SNFR)} problem.
It is important to recognize that the
recovery path depends heavily on the protocol being deployed in the
system.  Later on we discuss our (simple) routing protocol.

\subsection{Preliminaries}
\label{prelim}

Our communication network is modeled by an edge-weighted 
biconnected undirected 
graph $G=(V,E)$, with $n=|V|$ and $m=|E|$.
Each edge $e\in E$ has an 
associated cost (weight), denoted by $cost(e)$, which 
is a non-negative real number.
$p_G(s,t)$ denotes a shortest path between $s$ and $t$
in graph $G$ and $d_G(s,t)$ to denote its cost (weight). 

A shortest path tree $\mathcal{T}_s$ for a node 
$s$ is a collection of
$n-1$ edges $\{e_1,e_2,\ldots,e_{n-1}\}$ of $G$
which form a spanning tree of $G$ such
that the path from node $v$ to $s$ in $\mathcal{T}_s$
is a shortest path from $v$ to $s$ in $G$.
We say that $\mathcal{T}_s$ is rooted at node $s$. With
respect to this root we define the set of nodes that are
the {\em children} of each node $x$ as follows.
In $\mathcal{T}_s$ we say that every node $y$ 
that is adjacent to $x$ such
that $x$ is on the path in $\mathcal{T}_s$ from 
$y$ to $s$, 
is a child of $x$. For each node $x$ in the shortest 
paths tree, $k_x$ denotes the number of 
children of $x$ in the tree, and 
$\mathcal{C}_x = \{x_1, x_2, \ldots x_{k_x}\}$ 
denotes this set of children of the node $x$. 
Also, $x$ is said to be the {\em parent} of each 
$x_i\in \mathcal{C}_x$ in the tree $\mathcal{T}_s$.
With respect to $s$, the parent node, $p$, 
of a node $c$ is sometimes 
referred to as the {\em primary neighbor} or 
{\em primary router} of $c$, while $c$ is
referred to as an {\em upstream neighbor} or
{\em upstream router} of $p$. The children of
a particular node are said to be {\em siblings}
of each other.
$V_x(\mathcal{T})$ denotes the set of nodes 
in the subtree of $x$ in the tree $\mathcal{T}$ 
and $E_x\subseteq E$ denotes the set of all 
edges incident on the node $x$ in the graph $G$. 
We use $nextHop(x,y)$ to denote the next node from 
$x$ on the shortest path tree from $x$ to $y$.
Note that by definition, $nextHop(x,y)$ is 
the parent of $x$ in $\mathcal{T}_y$.

Finally, we use $\rho_x$ to denote the
escape edge in $G(E) \backslash \mathcal{T}_s$ 
that the node $x$ uses
to recover from the failure of its parent. As
we discuss later, having the information of a
single escape edge $\rho_x$ for each node
$x \in G(V)$ and $x \neq s$ is sufficient to 
construct the entire alternate path for any node
to recover from the failure of its parent, even
though the path may actually contain multiple
non-tree edges.

\subsection{Related Work}

One popular approach of tackling the issues related to transient
failures of network elements is that of using {\em proactive recovery
schemes}. These schemes typically work by precomputing alternate
paths at the network setup time for the failure scenarios, 
and then using these alternate
paths to re-route the traffic when the failure actually occurs.
Also, the information of the failure is suppressed
in the hope that it is a transient failure. The local 
rerouting based solutions proposed in 
\cite{bg,lynzc,sl,wg08,znylwc} fall into this category.

Refs. \cite{fir-ton,znylwc} present protocols based on 
local re-routing for dealing with transient single link 
and single node failures respectively. 
They demonstrate via simulations that the recovery paths computed 
by their algorithm are usually within 15\%
of the theoretically optimal alternate paths. 

Wang and Gao's Backup Route Aware Protocol \cite{wg08} also
uses some precomputed backup routes in order to handle transient
single {\em link} failures. One problem central to their 
solution asks for the availability of {\em reverse paths} 
at each node. However, they do not discuss the computation 
of these reverse paths. Interestingly, the alternate 
paths that our algorithm computes qualify as
the reverse paths required by the BRAP protocol 
of \cite{wg08}.

Slosiar and Latin \cite{sl} studied the single {\em link} failure
recovery problem and presented 
an $O(n^3)$ time for computing the link-avoiding alternate
paths. A faster algorithm, with a running time of $O(m + n\log n)$
for this problem was presented in \cite{bg}. Our central
protocol presented in this paper can be generalized to handle
single link failures as well. Unlike the protocol 
of \cite{fir-ton}, this single link failure recovery 
protocol would use {\em optimal} recovery paths.

\subsection{Problem Definition}
\label{def}
The Single Node Failure Recovery problem, is defined as follows:
({\tt SNFR}) Given a biconnected undirected edge weighted graph $G=(V,E)$,
and the shortest paths tree 
$\mathcal{T}_s(G)$ of a node $s$ in $G$ where 
$\mathcal{C}_x = \{x_1, x_2, \ldots x_{k_x}\}$
denotes the set of {\em children} of the node $x$ in 
$\mathcal{T}_s$, for each node $x \in V$ and $x \neq s$, find 
a path from $x_i \in \mathcal{C}_x$ to $s$ in the graph 
$G=(V\setminus \{x\}, E\setminus E_x)$, where $E_x$ is the
set of edges adjacent to vertex $x$.

In other words, for each node $x$ in the graph, we are interested 
in finding alternate paths from each of its children to the source\footnote{
We use {\em source} and {\em destination} in an interchangeable way}
node $s$ when the node $x$ {\em fails}. Note that
we don't consider the problem to be well defined 
when the node $s$ fails.

The above definition of alternate paths matches that in \cite{wg08}
for {\em reverse paths}: for each node $x\in G(V)$,
find a path from $x$ to the node $s$ that does not use 
the primary neighbor (parent node) $y$ of $x$ in $\mathcal{T}_s$.

\subsection{Main Results}
\label{main-results}

We discuss our efficient\footnote{The primary routing tables can be 
computed using the Fibonacci heaps \cite{fib} based implementation 
of Dijkstra's shortest paths algorithm \cite{dijk} in $O(m + n\log n)$ 
time} algorithm for the SNFR problem that has a running time 
of $O(m \log n)$ (by contrast, the
alternate path algorithms of \cite{lynzc,fir-ton,znylwc} 
have a time complexity 
of $\Omega(mn\log n)$ per destination).
We further develop protocols based on this algorithm 
for recovering from single node {\em transient} failures in 
communication networks. In the {\em failure free} case,
our protocol does not use any extra resources.

The recovery paths computed by our algorithm are not 
necessarily the shortest recovery paths. However,
we demonstrate via simulation results that they
are very close to the optimal paths.
  
We compare our results with those of \cite{znylwc} 
wherein the authors have also studied the same problem and 
presented protocols based on local rerouting for dealing 
with transient single node failures. One important
difference between the algorithms of \cite{lynzc,fir-ton,znylwc}
and our's is that unlike our algorithm, these 
are based primarily on recomputations.
Consequently, our 
algorithm is faster by an order of magnitude than 
those in \cite{lynzc,fir-ton,znylwc}, and as shown by our 
simulation results, our recovery paths are usually 
comparable, and sometimes better.

\section{Algorithm for Single Node Failure Recovery}
\label{snfr}

A naive algorithm for the SNFR problem is based on recomputation: 
for each node $v \in G(V)$ and $v\neq s$, compute the shortest 
paths tree of $s$ in the graph $G(V\backslash v, E\backslash E_v)$. 
Of interest are the paths from $s$ to each of the nodes 
$v_i\in \mathcal{C}_v$. This naive algorithm invokes a shortest
paths algorithm $n-1$ times, and thus takes $O(mn + n^2\log n)$ 
time when it uses the Fibonacci heap \cite{fib} implementation 
of Dijkstra's shortest paths algorithm \cite{dijk}. While these 
paths are {\em optimal} recovery paths for recovering from the 
node failure, their {\em structure} can be much different 
from each other, and from the original shortest paths (in absence 
of any failures) - to the extent that routing messages along these 
paths may involve recomputing large parts of the primary routing 
tables at the nodes through which these paths pass. 
The recovery paths computed by our algorithm have a well 
defined structure, and they overlap with the paths in the 
original shortest paths tree ($\mathcal{T}_s$) to an extent that
storing the information of a single edge, $\rho_x$,
at each node $x$ provides 
sufficient information to infer the entire recovery path.

\subsection{Basic Principles and Observations}
\label{sub-graph}

We start by describing some basic observations about the
characteristics of the recovery paths.
We also categorize the graph edges according to their {\em role} in 
providing recovery paths for a node when its parent fails. 

\begin{figure}[h]
\centerline{\epsfig{file=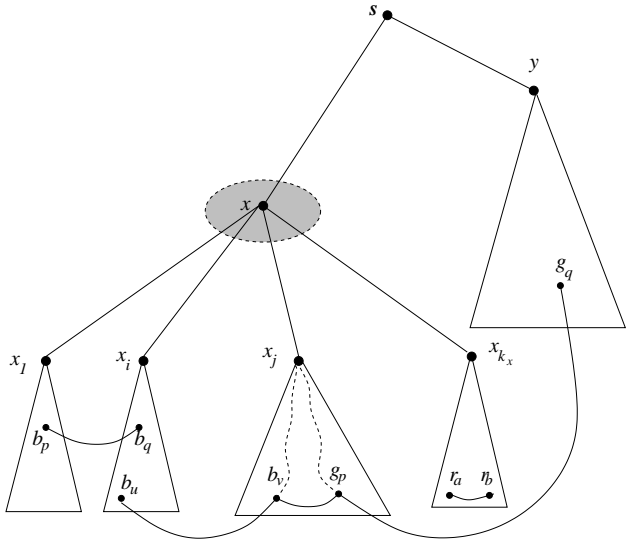 , width = 1.0\linewidth}}
\caption{Recovery paths for recovering from the failure of $x$.}
\label{basic}
\end{figure}

Figure \ref{basic} illustrates a scenario of a single node failure. 
In this case, the node $x$ has failed, and we need to find recovery 
paths to $s$ from each $x_i\in \mathcal{C}_x$. When a node fails, 
the shortest paths tree of $s$, $\mathcal{T}_s$, gets split into 
$k_x + 1$ components - one containing the source node $s$ and each 
of the remaining ones contain one subtree of a child 
$x_i \in \mathcal{C}_x$.

Notice that the edge $\{g_p, g_q\}$ (Figure \ref{basic}),
which has one end point in the subtree of $x_j$, and the other 
outside the subtree of $x$ provides a candidate recovery path 
for the node $x_j$. The complete path is of the form 
$p_G(x_j, g_p) \leadsto \{g_p, g_q\} \leadsto p_G(g_q, s)$. 
Since $g_q$ is outside the subtree of $x$, the path $p_G(g_q, s)$ 
is not affected by the failure of $x$. Edges of this type 
(from a node in the subtree of $x_j \in \mathcal{C}_x$ to a 
node outside the subtree of $x$) can be used by 
$x_j \in \mathcal{C}_x$ to {\em escape} the failure of node 
$x$. Such edges are called {\em green} edges.  For example,
edge $\{g_p, g_q\}$ is a green edge.

Next, consider the edge $\{b_u, b_v\}$ (Figure \ref{basic})
between a node in the subtree of $x_i$ and a node in the subtree 
of $x_j$. Although there is no green edge with an end point in 
the subtree of $x_i$, the edges $\{b_u, b_v\}$ and $\{g_p, g_q\}$ 
together offer a candidate recovery path that can be used by 
$x_i$ to recover from the failure of $x$. Part of this path 
connects $x_i$ to $x_j$ 
($p_G(x_i, b_u) \leadsto \{b_u, b_v\} \leadsto p_G(b_v, x_j)$), 
after which it uses the recovery path of $x_j$ (via $x_j$'s 
green edge, $\{g_p, g_q\}$). Edges of this type (from a node 
in the subtree of $x_i$ to a node in 
the subtree of a sibling $x_j$ for 
some $i \not= j$) are called {\em blue} edges. Another example 
of a blue edge is edge $\{b_p, b_q\}$ which can be used the 
node $x_1$ to recover from the failure of $x$.

Note that edges like $\{r_a, r_b\}$ and $\{b_v, g_p\}$ (Figure 
\ref{basic}) with both end points within the subtree of the
same child of $x$ do not help any of the nodes in $\mathcal{C}_x$
to find a recovery path from the failure of node $x$.
We do not consider such edges in the computation of recovery paths,
even though they may provide a shorter recovery path for some nodes
(e.g. $\{b_v,g_p\}$ may offer a shorter recovery path to $x_i$).
The reason for this is that routing protocols would need to be quite
complex in order to use this information.
We carefully organize the {\em green} and {\em blue} 
edges in a way that allows us to retain only the useful edges and 
eliminate useless (red) ones efficiently.

We now describe the construction of a new graph $\mathcal{R}_x$,
the recovery graph for $x$, which
will be used to compute recovery paths for the elements of 
$\mathcal{C}_x$ when the node $x$ fails.
A single source shortest paths computation on this graph 
suffices to compute the recovery paths for all 
$x_i \in \mathcal{C}_x$.

The graph $\mathcal{R}_x$ has $k_x + 1$ nodes, where $k_x = |\mathcal{C}_x|$. 
A special node, $s_x$, represents the source node $s$ in the 
original graph $G=(V,E)$. Apart from $s_x$, we have
one node, denoted by $y_i$, for each $x_i\in \mathcal{C}_x$.
We add all the {\em green} and {\em blue} edges defined earlier 
to the graph $\mathcal{R}_x$ as follows. A green edge with an end point 
in the subtree of $x_i$ (by definition, green edges have the 
other end point outside the subtree of $x$) translates 
to an edge between $s_x$ and $y_i$. A blue edge with an end 
point in the subtree of $x_i$ and the other in the subtree of 
$x_j$ translates to an edge between nodes $y_i$ and $y_j$.
However, the weight of each edge added to $\mathcal{R}_x$ is not the 
same as the weight of the green or blue edge in $G=(V,E)$ 
used to define it. The weights are specified 
below.

Note that the candidate recovery path of $x_j$ that uses the 
green edge $g = \{g_p, g_q\}$ has total cost equal to:

\begin{equation}
\label{green-weight}
greenWeight(g) = d_G(x_j, g_p) + cost(g_p, g_q) + d_G(g_q, s)
\end{equation}

As discussed earlier, a blue edge provides a path connecting
two siblings of $x$, say $x_i$ and $x_j$. Once the path reaches 
$x_j$, the remaining part of the recovery path of $x_i$ 
coincides with that of $x_j$. If $\{b_u, b_v\}$ is the blue 
edge connecting the subtrees of $x_i$ and $x_j$ (the cheapest 
one corresponding to the edge $\{y_i,y_j\}$), the length
of the subpath from $x_i$ to $x_j$ is:
\begin{equation}
\label{blue-weight}
blueWeight(b) = d_G(x_i, b_u) + cost(b_u, b_v) + d_G(b_v, x_j)
\end{equation}

We assign this weight to the edge corresponding to the blue edge
$\{b_u, b_v\}$ that is added 
in $\mathcal{R}_x$ between $y_i$ and $y_j$.

The construction of our graph $\mathcal{R}_x$ is now complete.
Computing the shortest paths tree of $s_x$ in $\mathcal{R}_x$ provides 
enough information to compute the recovery paths for all nodes 
$x_i \in \mathcal{C}_x$ when $x$ fails.

\subsection{Description of the Algorithm and its Analysis}

We now incorporate the basic observations described earlier into a 
formal algorithm for the SNFR problem. Then we analyze the complexity 
of our algorithm and show that it has a nearly optimal running time 
of $O(m\log n)$. 

Our algorithm is a {\em depth-first} recursive algorithm over 
$\mathcal{T}_s$. We maintain the following information at each node $x$:
\begin{itemize}
\item Green Edges: The set of green edges in $G=(V,E)$ that offer a 
recovery path for $x$ to escape the failure of its parent.
\item Blue Edges: A set of edges $\{p,q\}$ in $G=(V,E)$ such that 
$x$ is the nearest-common-ancestor of $p$ and $q$ with respect to 
the tree $\mathcal{T}_s$.
\end{itemize}

The set of green edges for node $x$ is maintained in a {\em min heap}
({\em priority queue}) data structure, which is denoted by
$\mathcal{H}_x$. The heap elements are tuples of the form
$<e, greenWeight(e)+d_G(s,x)>$ where $e$ 
is a green edge, and $greenWeight(\cdot)+d_G(s,x)$
defines its priority as an element of the heap.
Note that the extra element $d_G(s,x)$ is added in order to maintain
invariance that the priority of an edge in any heap $\mathcal{H}$ 
remains constant as the path to $s$ is traversed.
Initially $\mathcal{H}_x$ contains an entry for each edge of $x$ 
which serves as a green edge for it (i.e. an edge of $x$ whose 
other end point does not lie in the subtree of the parent of $x$). 
A linked list, $\mathcal{B}_x$, stores the tuples $<e, blueWeight(e)>$, 
where $e$ is a blue edge, and $blueWeight(e)$ is the weight of
$e$ as defined by the equation (\ref{blue-weight}).

The heap $\mathcal{H}_{x_i}$ is built by merging 
together the $\mathcal{H}$ heaps of the nodes in $\mathcal{C}_{x_i}$, 
the set of children on $x_i$. Consequently, all the elements in 
$\mathcal{H}_{x_i}$ may not be green edges for $x_i$. Using a
dfs labeling scheme similar to the one in \cite{bg}, 
we can quickly determine whether the edge 
retrieved by $findMin(\mathcal{H}_{x_i})$ is a valid green edge 
for $x_i$ or not. If not, we remove the entry corresponding to 
the edge from $\mathcal{H}_{x_i}$ via a $deleteMin(\mathcal{H}_{x_i})$ 
operation. Note that since the deleted edge cannot serve as a 
green edge for $x_i$, it cannot serve as one for any of the 
ancestors of $x_i$, and it doesn't need to be added back to the 
$\mathcal{H}_x$ heap for any $x$. We continue deleting the minimum 
weight edges from $\mathcal{H}_{x_i}$ till either $\mathcal{H}_{x_i}$ 
becomes empty or we find a green edge valid for $x_i$ to escape 
$x$'s failure, in which case we add it to $\mathcal{R}_x$.  

After adding the green edges to $\mathcal{R}_x$, we add the blue edges from 
$\mathcal{B}_x$ to $\mathcal{R}_x$. 

Finally, we compute the shortest paths tree of the node $s_x$ in 
the graph $\mathcal{R}_x$ using a standard shortest paths algorithm 
(e.g. Dijkstra's algorithm \cite{dijk}). The {\em escape edge} 
for the node $x_i$ is 
stored as the {\em parent edge} of $x_i$ in $\mathcal{T}_{s_x}$, 
the shortest paths tree of $s_x$ in $\mathcal{R}_x$.
Since the communication graph is assumed to be {\em bi-connected}, 
there exists a path from each node $x_i\in \mathcal{C}_x$ to $s_x$, 
provided that the failing node is not $s$.

For brevity, we omit the detailed analysis of the algorithm. The
$O(m\log n)$ time complexity of the algorithm follows from the fact
that (1) An edge can be a blue edge in the recovery graph of exactly one
node: that of the nearest-common-ancestor of its two end points, and
(2) An edge can be deleted at most once from any $\mathcal{H}$ heap.
We state the result as the following theorem.

\begin{theorem}
Given an undirected weighted graph $G=(V,E)$ and a specified node 
$s$, the recovery path from each node $x_i$ to $s$ to escape from 
the failure of the parent of $x$ is computed by our procedure in 
$O(m\log n)$ time.
\end{theorem}

\section{Single Node Failure Recovery Protocol}
\label{protocol}

When routing a message to a node $s$, if a node $x$ needs to 
forward the message to another node $y$, the node $y$ is the {\em parent} 
of $x$ in the shortest paths tree $\mathcal{T}_s$ of $s$. The SNFR 
algorithm computes the recovery path from $x$ to $s$ which does not 
use the node $y$. In case a node has failed, the protocol 
re-routes the messages along these alternate paths that have been 
computed by the SNFR algorithm.

\subsection{Embedding the Escape Edge}
In our protocol, the node $x$ that discovers the failure 
of $y$ embeds information about the escape edge to use in the message. 
The escape edge is same as the $\rho_x$ edge 
identified for the node $x$ 
to use when its parent ($y$, in this example) has failed. 
We describe two alteratives for embedding the escape edge information
in the message, depending on the particular routing protocol being
used.

\noindent
\newline
{\bf Protocol Headers}

In several routing protocols, including TCP, the message headers
are not of fixed size, and other header fields (e.g. 
{\tt Data Offset} in TCP) indicate where the actual 
message data begins. For our purpose, we need an additional header 
space for two node identifiers (e.g. IP addresses, and the port 
numbers) which define the two end points of the escape edge. 
It is important to note that this extra space is required only when 
the messages are being re-routed as part of a failure 
recovery. In absence of failures, we do not need to modify the 
message headers.

\noindent
{\bf Recovery Message}

In some cases, it may not be feasible or desirable to add the information
about the escape edge to the protocol headers. In such situations, the
node $x$ that discovers the failure of its parent node $y$ during the
delivery of a message $\mathcal{M}_o$, constructs a new message, 
$\mathcal{M}_r$, that contains information for recovering from the
failure. In particular, the recovery message, $\mathcal{M}_r$ contains
(a) $\mathcal{M}_o$: the original message, and (b) $\rho_x = 
(p_x, q_x)$: the escape edge to be used by $x$ to recover from the 
failure of its parent.

With either of the above two approaches, a light weight application
is used to determine if a message is being routed in a {\em failure
free} case or as part of a {\em failure recovery}, and take appropriate
actions. Depending on whether the escape edge information is present
in the messagae, the application decides which node to forward the
message to. This process consumes almost negligible additional resources.
As a further optimization, this application can 
use a special reserved port
on the routers, and messages would be sent to it only during the
failure recovery mode. This would ensure that no additional
resources are consumed in the failure free case.

\subsection{Protocol Illustration}
For brevity we do not formally specify our protocol, but only 
illustrate how it works. Consider the network in Figure \ref{basic}.
If $x_i$ notices that $x$ has failed, it adds information in
the message (using one of the two options discussed above) about 
$\{b_u, b_v\}$ as the escape edge to 
use, and reroutes the message to $b_u$. $b_u$ clears the escape
edge information, and sends the message to $b_v$, after which 
it follows the {\em regular} path to $s$. If $x$ has not 
recovered when
the message reaches $x_j$, $x_j$ reroutes with message to $g_p$
with $\{g_p,g_q\}$ as the escape edge to use. This continues
till the message reaches a node outside the subtree of $x$, or
till $x$ recovers.

Note that since the alternate paths are used only during
failure recovery, and the escape edges dictate the alternate
paths, the protocol ensures {\em loop free} routing, even though
the alternate paths may form loops with the original routing
(shortest) paths.

\section{Simulation Results and Comparisons}
\label{simu}

We present the simulation results for our algorithm, 
and compare the lengths of the recovery paths generated by our 
algorithm to the theoretically optimal paths as well as with
the ones computed by the algorithm in \cite{znylwc}. 
In the implementation of our algorithm, 
we have used {\em standard} data structures (e.g. binary heaps 
instead of Fibonacci heaps \cite{fib}: binary heaps suffer from a 
linear-time merge/meld operation as opposed to constant time for 
the latter). Consequently, our algorithms have the potential to 
produce much better running times than what we report. 

\begin{figure}[h]
\centerline{\epsfig{file=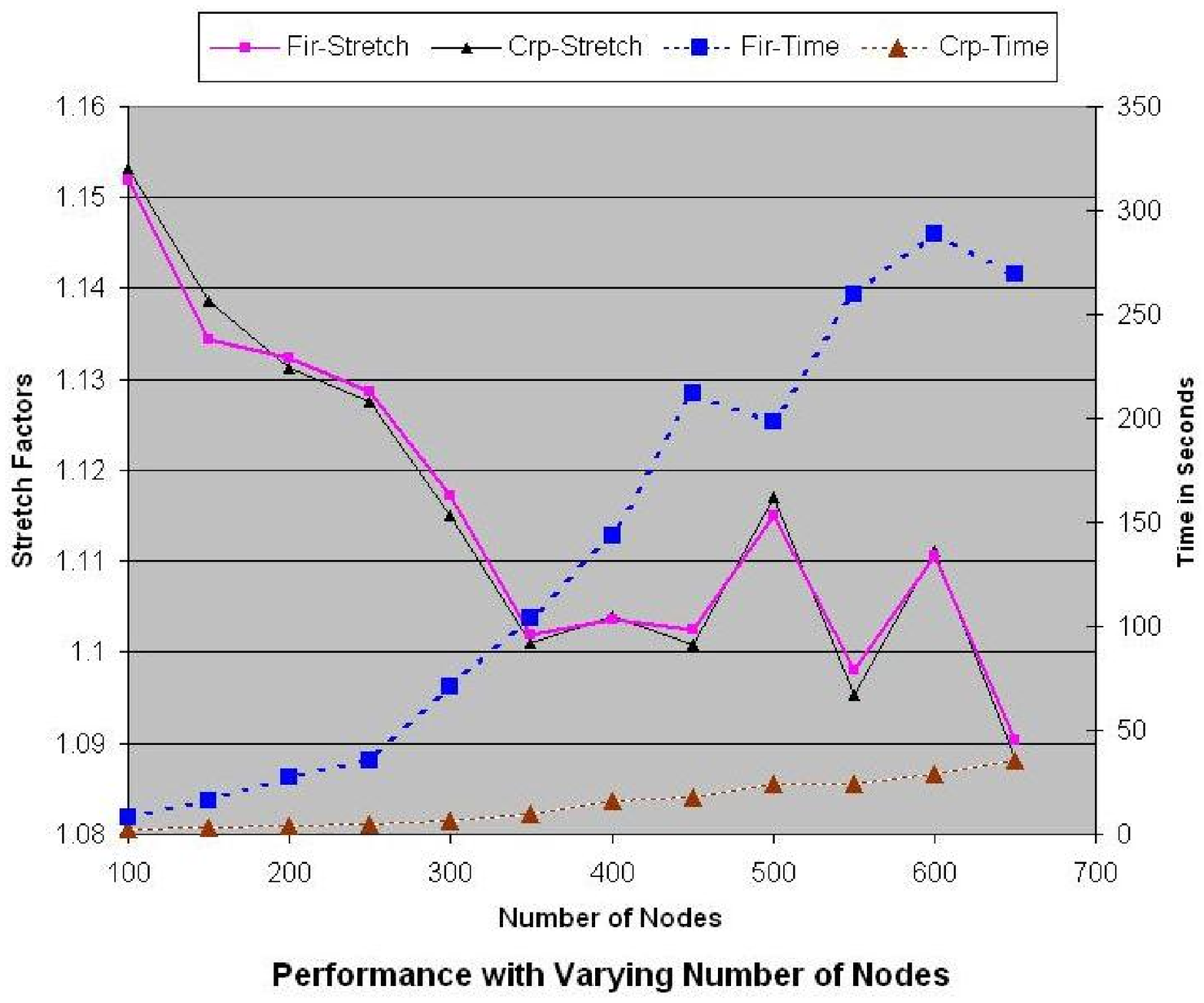, width = 1.1\linewidth}}
\caption{}
\label{compare-nodes}
\end{figure}

We ran our simulations on randomly generated graphs, with varying 
the following parameters: $(a)$ Number of nodes, and $(b)$ Average 
degree of a node. The edge weights are randomly generated numbers 
between 100 and 1000. In order to guarantee that the graph is 
{\em 2-node-connected} (biconnected), we ensure that the generated 
graph contains a {\em Hamiltonian cycle}.
Finally, for each set of these parameters, 
we simulate our algorithm on multiple random graphs to compute the 
{\em average} value of the of a {\em metric} for the parameter set.
The algorithms have been implemented in the Java
programming language (1.5.0.12 patch),
and were run on an Intel 
machine (Pentium IV 3.06GHz with 2GB RAM).

\begin{figure}[h]
\centerline{\epsfig{file=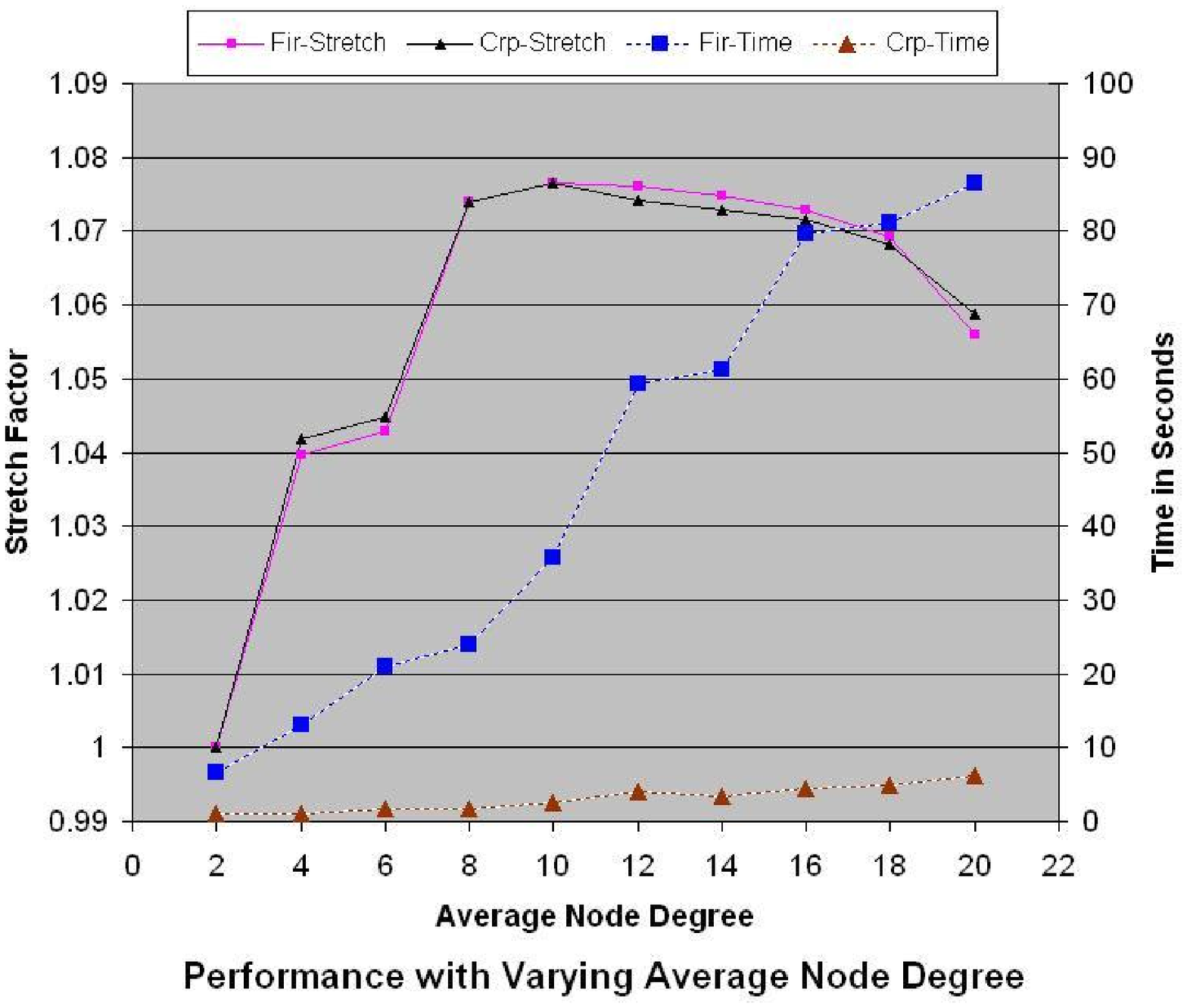, width = 1.1\linewidth}}
\caption{}
\label{compare-degree}
\end{figure}

The {\em stretch factor} is defined as the ratio of the lengths of recovery 
paths generated by our algorithm to the lengths of the theoretically 
optimal paths. The optimal recovery path lengths
are computed by recomputing the shortest paths tree of $s$ 
in the graph $G(V\backslash x, E\backslash E_x)$.
In the figures [\ref{compare-nodes},\ref{compare-degree}], the {\tt Fir}
labels relate to the performance of the alternate paths algorithm used by
the Failure Insensitive Routing protocol of \cite{znylwc}, while the
{\tt Crp} labels relate to the performance of our algorithm for the
SNFR problem.

Though \cite{znylwc} doesn't present a detailed analysis of 
their algorithm, from our analysis, their 
algorithm needs at least $\Omega(mn\log n)$ time
{\em per sink node} in the system. 
Figures [\ref{compare-nodes},\ref{compare-degree}] compare the performance 
of our algorithm ({\tt CRP}) to that of \cite{znylwc} ({\tt FIR}).
The plots for the running times of our 
algorithm and that of \cite{znylwc} fall in line with the 
theoretical analysis that our algorithms are faster by an order of 
magnitude than those of \cite{znylwc}. Interestingly, the stretch 
factors of the two algorithms are very close for most of the cases,
and stay within 15\%. The running time of the algorithms fall in
line with our theoretical analysis. Our {\tt CRP} algorithm runs
within 50 seconds for graphs upto 600-700 nodes, while the {\tt FIR}
algorithm's runtime shoots up to as high as 5 minutes as the
number of nodes increase.
The metrics are plotted against 
the variation in (1) the number of nodes 
(Figure [\ref{compare-nodes}]), and (2) the average 
degree of the nodes (Figure [\ref{compare-degree}]). 
The average degree of a 
node is fixed at $15$ for the cases where we vary the number of nodes 
(Figure [\ref{compare-nodes}]), and the number of nodes is fixed
at $300$ for the cases where we plot the impact of varying average node
degree (Figure [\ref{compare-degree}]).
As expected, the stretch factors {\em improve} as the number of 
nodes increase. Our algorithm falls behind in finding the optimal 
paths in cases when the recovery path passes through the subtrees 
of multiple siblings. Instead of finding the best {\em exit} 
point out of the subtree, in order to keep the protocol simple 
and the paths well {\em structured}, our paths go to the root 
of the subtree and then follow its alternate path beyond that. 
These paths are formed using the blue edges. Paths discovered 
using a node's green edges are optimal such paths. In other 
words, if most of the edges of a node are green, our algorithm 
is more likely to find paths close to the optimal ones. Since
the average degree of the nodes is kept fixed in these 
simulations, increasing the number of nodes increases the 
probability of the edges being green. A similar logic explains
the plots in Figure [\ref{compare-degree}]. When
the number of nodes is fixed, increasing the average degree of
a node results in an increase in the number of green edges
for the nodes,\footnote{When the average degree is very small,
there are only a few alternate paths available, and the 
algorithms usually find the better ones among them, 
resulting in smaller stretch factors.}
as well as the stretch factors.

\section{Concluding Remarks}
\label{concl}

In this paper we have presented an efficient algorithm for
the SNFR problem, and developed protocols for dealing with
transient single node failures in communication networks. 
Via simulation results, we show that our algorithms are 
much faster than those of \cite{znylwc}, while the stretch 
factor of our paths are usually better or comparable.

Previous algorithms \cite{lynzc,fir-ton,znylwc} for 
computing alternate paths are much slower, and thus
impose a much longer network setup time as
compared to our approach. The setup time becomes
critical in more dynamic networks, where the
configuration changes due to events other than
transient node or link failures. Note that in several 
kinds of configuration changes (e.g. permanent 
node failure, node additions, etc), recomputing the
routing paths (or other information) cannot be
avoided, and it is desirable to have shorter 
network setup times.

For the case where we need to solve the SNFR problem for 
{\em all} nodes in the graph, our algorithm would need
$O(mn\log n)$ time, which is still very close to the
time required ($O(mn+n^2\log n)$) to build the routing 
tables for the all-pairs setting. The space requirement
still stays linear in $m$ and $n$.

The {\em directed} version of the SNFR problem, where one needs to 
find the {\em optimal} (shortest) recovery paths can be shown to have 
a lower bound of $\Omega(min(m\sqrt{n},n^2))$ using a construction 
similar to those used for proving the same lower bound on the directed 
version of SLFR\cite{bg} and replacement paths\cite{hsb} problems. 
The bound holds under the {\em path comparison} model of \cite{kkp} 
for shortest paths algorithms.

\small
\bibliography{snfr-corr}
\bibliographystyle{plain}

\end{document}